\documentclass[aps,prd,nofootinbib,onecolumn,groupedaddress,amsfonts,amssymb,amsmath,showpacs,showkeys]{revtex4-1}
\usepackage[backref,breaklinks,colorlinks,citecolor=blue]{hyperref}
\usepackage{mathtools}

\usepackage{graphicx}

\usepackage{color,soul}

\begin{document}

\title{A direct primitive variable recovery scheme for hyperbolic
conservative equations: the case of relativistic hydrodynamics}

\author{A. Aguayo-Ortiz$^{1}$}
\email[Email address:]{aaguayo@astro.unam.mx}
\author{S. Mendoza$^{1}$}
\email[Email address: ]{sergio@astro.unam.mx}
\author{D. Olvera$^{2}$}
\email[Email address: ]{do12542@bristol.ac.uk}
\affiliation{$^1$Instituto de Astronom\'{\i}a, Universidad Nacional
                 Aut\'onoma de M\'exico, AP 70-264, Ciudad de M\'exico 04510,
	         M\'exico \\
             \(^2\) School of Mathematics, University of Bristol,  Bristol
	            BS8 1TW,  United Kingdom
            }

\date{\today}


\begin{abstract}
  {In this article we develop a Primitive Variable Recovery Scheme (PVRS)
  to solve any system of coupled differential conservative equations. This
  method obtains directly the primitive variables applying the chain
  rule to the time term of the conservative equations.  With this, a
  traditional finite volume method for the flux is applied in order avoid
  violation of both, the entropy and ``Rankine-Hugoniot'' jump conditions.
  The time evolution is then computed using a forward finite difference
  scheme. This numerical technique evades the recovery of the primitive
  vector by solving an algebraic system of equations as it is often used and
  so, it generalises standard techniques to solve these kind of coupled
  systems.  The article is presented bearing in mind special relativistic
  hydrodynamic numerical schemes with an added pedagogical view in the
  appendix section in order to easily comprehend the PVRS.  We present
  the convergence of the method for standard shock-tube problems of
  special relativistic hydrodynamics and a graphical visualisation of
  the errors using the fluctuations of the numerical values with respect
  to exact analytic solutions.
  The PVRS circumvents the sometimes arduous computation that 
  arises from standard numerical methods techniques, which obtain the desired
  primitive vector solution through an algebraic polynomial of the
  charges.  }
\end{abstract}

\maketitle

\section{Introduction}
\label{introduction}

 The use of numerical methods to solve differential equations has
constituted a substantial amount of work since the conception of
approximate solutions to a given set of equations.  In the last few
decades, digital computers have been a great help to heavily iterate
complicated partial differential equations using extensive numerical,
parallel and adaptive mesh techniques in personal computers and large
clusters.

 Physical laws are often written in a set of conservative differential
equations, for which there are many well established convergent numerical
techniques to obtain accurate solutions.  In spite of this, there is an
intermediate step that is often, depending on the nature of the problem,
extremely cumbersome to deal with.  This appears since the general solution
to the problem is obtained as a set of vector charges \( \boldsymbol{q} \)
at every point or cell on a given domain of space at a particular time in
the iteration.  However, physical phenomena are described and measured by
means of a set of vector primitive variables \( \boldsymbol{u} \).
Depending on the nature of the physical problem, the function \(
\boldsymbol{u}(\boldsymbol{q}) \) may not have an analytic form and so,
at every point or cell of the integration space a cumbersome technique
requires to be performed for each time step.  No matter how fast this
routine may be, it introduces an extra computational time that can heavily
grow when the space-time resolution increases. In
problems of special relativistic hydrodynamics this fact appears and, at
each time step, a 10th degree algebraic polynomial has to be solved for a
unique given value of each component of the vector \( \boldsymbol{u}
\)~\citep[see e.g.][for an excellent account on this]{riccardi2008}.

  To make things even more complicated, for each particular physical
problem it is necessary to have either an analytic solution \(
\boldsymbol{u}(\boldsymbol{q}) \) or a specific numerical technique to
obtain it.  

  In this article we show how it is possible to construct a general
numerical iteration method, using {a combination of finite differences and 
finite volume integration techniques for the time and spatial evolutions
respectively,} to directly find the solutions \( \boldsymbol{u}
\) avoiding any middle cumbersome step such as the ones mentioned above.
This technique is so general that requires no {analytical} knowledge 
whatsoever of \(
\boldsymbol{u}(\boldsymbol{q}) \).  The method developed is general and 
valid to any set of coupled conservative equations.  We also show how
this method can be applied in the particular case of 
1D special relativistic hydrodynamics (1DRHD).  For this
particular case, we construct convergence tests.

  The article is organised as follows.  {In the appendix}
section~\ref{traditional}, 
we briefly mention some (mostly used in relativistic hydrodynamics
for shock capturing) of the traditional methods to solve a set of
conservative equations.  In section~\ref{primitive} we construct our
``\emph{Primitive Variable Recovery Scheme} (PVRS)'' which can directly
obtain the primitive variables from quite a standard numerical procedure.
Section~\ref{convergence} deals with different {convergence relativistic 
Sod \protect{\citep{sod}} shock-tube tests}
and error estimates are given using a standard L$_1$-norm. Also, {the errors
are graphically interpreted} using the fluctuations
of the solution with respect to analytical known values is presented.  Finally,
in section~\ref{discussion} we discuss and conclude our results.

\section{Primitive variable recovery scheme (PVRS)}
\label{primitive}

{  In the appendix we discuss some of the standard techniques for
discretising any set of scalar and coupled conservative equations.
This is done in order to easy understand the further developments of the
article for the less expert reader, and not to interrupt the experienced
one with such well known methods. However, we note that in the appendix and
in what follows Einstein's summation convention will be used throughout the
equations displayed in this article, something that does not usually appear
in the literature.}

{  The usual way to solve a system of hyperbolic equations (cf.
equation~\protect{\eqref{eq:conservative}):}}
  
\begin{equation}\label{eq:conservative-vector}
{\protect{  \frac{\partial \boldsymbol{q} }{ \partial t} + \frac{\partial
  \boldsymbol{f}(\boldsymbol{q})}{\partial x} = 0,}}
\end{equation}

\noindent  {is by implementing Finite Difference and Finite Volume
Methods (FDM \& FVM) in order to obtain solutions for the conservative
charges $\boldsymbol{q}$.} In the particular case of relativistic and
non-relativistic hydrodynamics, these charges are the linear momentum
along the three dimensions $S^i$, the energy $\tau$ and the particle
density $D$. In order to compare the numerical solution with
experiments and/or observations, a set of primitive physical measurable
variables $u$ needs to be constructed.  For this particular case, this
primitive variable set is given by by the pressure $p$, the velocity
along three spacial dimensions $v^i$ and, the particle number density
$n$\footnote{Some authors prefer to find the particle mass density \(
\rho \) rather than the particle number density \( n \).  For most
practical proposes, both variables are related by $\rho = mn$ where $m$
is the average mass per particle.}.  {In here and in what follows all
thermodynamical quantities (pressure $p$, particle number density $n$ and
energy density $e$ and so on, are measured on its proper reference frame
following the convention by \protect{\citep{landau1987,weinberg}})}.  The
explicit dependences $\boldsymbol{q}=\boldsymbol{q}(\boldsymbol{u}(x,t))$
and $\boldsymbol{f}=\boldsymbol{f}(\boldsymbol{u}(x,t))$ for
1D flow  in the special relativistic case are given by (see
e.g.~\citep{weinberg,landau1987}):

\begin{gather}
q_1 = D = \frac{n}{\sqrt{1-v^2}} \quad \textup{and} \quad	f_1 = v\frac{n}{\sqrt{1 - v^2}},
\label{eq:q1} \\
q_2 = S^x = v \frac{e + p}{1 - v^2} \quad \textup{and} \quad f_2 = v^2\frac{e+p}{1-v^2} +p ,
\label{eq:q2} \\
q_3 = \tau = \frac{e + v^2 p}{1 - v^2} \quad \textup{and} \quad f_3 = v\frac{e+p}{1-v^2}.
\label{eq:q3}
\end{gather}

\noindent where $e$ is the total (rest plus internal) proper energy per
unit volume which can be related with the density and pressure via
a state equation $e=e(n,p)$ like the one derived by~\citep{tooper1965}
for a polytropic relativistic gas:

\begin{equation}
  e = nm + \frac{ p }{ \kappa - 1 }, 
\end{equation}

\noindent where \( \kappa \) is the polytropic index.  In the previous
equations and in what follows we {choose} a system of units in which the
velocity of light is set to unity.

As we can see from relations~(\ref{eq:q1})-(\ref{eq:q3}), obtaining the
inverse function $\boldsymbol{u}=\boldsymbol{u}(\boldsymbol{q}(x,t))$ results in quite a completed algebraic
problem.  In fact, the solution to this problem leads to a system
of transcendental algebraic equations that have been deeply studied
by~\citep{riccardi2008}. One way of solving this system is by using a
Newton-Raphson method \citep[cf.][]{yousaf2015} but this or any other numerical
solution to obtain $\boldsymbol{u}(\boldsymbol{q}(x,t))$ will carry an extra error besides the proper
numerical error of the FDM or FVM.  This procedure also adds a bit of
computational processing time since an iteration loop to find the solution
needs to be carried out at each cell every time step.  In order to avoid
this cumbersome task, we show now how it is possible to obtain a direct
numerical solution of the primitive variables, which is valid for all
conservative equation systems~\eqref{eq:conservative}.

\section{PVRS attempts with Finite Difference Methods.}

\label{finite-diff-pvrs}

  Let us begin by writing  the system of $m$ hyperbolic equations
showing the explicit dependence on $m$ primitive variables, i.e.:

\begin{equation}\label{eq:prim-cons-eq}
\frac{\partial q_a(u_1,...,u_m)}{\partial t} + \frac{\partial f_a(u_1,...,u_m)}{\partial x} = 0.
\end{equation}

\noindent A necessary and sufficient condition for the existence of the
solution \( u_1,\ldots,u_m\) is that \( a=1,\ldots,m\).  Now, using
the chain rule, the above equation can be written in the following
quasilinear form:

\begin{equation}
\frac{\partial q_a}{\partial u_b}\frac{\partial u_b}{\partial t} + \frac{\partial f_a}{\partial u_c}\frac{\partial u_c}{\partial x} = 0,
\end{equation}

\noindent where $\partial \boldsymbol{q} / \partial \boldsymbol{u}$ and
$\partial \boldsymbol{f} / \partial \boldsymbol{u}$ are the 
Jacobian matrixes of the vectors \( \boldsymbol{q} \) and \( \boldsymbol{f}
\) respectively. Multiplying the previous equation by the inverse matrix
$( \partial \boldsymbol{q} / \partial \boldsymbol{u})^{-1} $ we get

\begin{equation}\label{eq:azt-qua-eq}
\frac{\partial u_a}{\partial t} + M_{ab} \frac{\partial u_b}{\partial x} = 0,
\end{equation}
\noindent where
\begin{equation}\label{eq:M}
M_{ab} := \left( \frac{\partial q_c}{\partial u_a} \right)^{-1} \left( \frac{\partial f_c}{\partial u_b} \right).
\end{equation}

If we {perform} a discretisation of equation~(\ref{eq:azt-qua-eq})
using a FDM {(see e.g. section~{\protect\ref{finite-diff}} of the appendix)},
we obtain the following numerical expression:

\begin{equation}\label{eq:M-equation}
\begin{split}
u_a(x_i,t_{n+1}) = u_a(x_i,t_n) - \frac{\Delta t}{2\Delta x} M_{ab}[&u_b(x_{i+1},t_n) \\
- &u_b(x_{i-1},t_n)].
\end{split}
\end{equation}

 No matter how complicated the  functional representations of 
\( \boldsymbol{q}(\boldsymbol{u}) \) and \(
\boldsymbol{f}(\boldsymbol{u}) \), it is possible (if not by hand, using a
Computer Algebra System) to compute the matrix \( M_{ab} \) only once
before implementing a discretisation scheme.  In what follows we show how
to implement a numerical scheme to find directly the primitive variables \(
\boldsymbol{u} \) solving equation~\eqref{eq:azt-qua-eq}. 
By doing this, the cumbersome step of recovering 
$\boldsymbol{u}$ from $\boldsymbol{q}$ at every cell for each time step
is not needed anymore.

 The discretisation~{\protect \eqref{eq:M-equation}} is accurate to the
first-order and
yields quite good results on smooth solutions. When the solution 
contains a shock wave, the method is stable 
but not consistent and so no convergent. This could be understood because
equation~({\protect\ref{eq:M-equation}}) is mathematically similar to 
relation~({\protect\ref{eq:quasilinear}}) of the
appendix~\citep{font1994} with the substitution of the vector
$ \boldsymbol{u} $ instead of \(
\boldsymbol{q} \). Furthermore,
equation~{\protect{\eqref{eq:M-equation}}} is written
in a non-conservative form and so, the entropy and Rankine-Hugoniot jump 
conditions are not satisfied across the shock waves. Due to this fact, the
obtained solution converges to a different weak solution as compared to the
one obtained by a conservative method (see
e.g.~{\protect{\citep{leveque2002}}}).  
In other words, this FDM scheme does not work  and the approach to follow
is to consider flux contributions as in standard FVM.

\section{Primitive Variable Recovery Scheme using combined FDM and FVM}
\label{finite-vol-pvrs}

{  We now show how to implement a Primitive Variable Recovery Scheme (PVRS)
using both a FDM and a FVM schemes for the time and spatial evolution of
the equations.  As mentioned at the end of the previous section,
the fluxes contribution in the method must not be altered because the
entropy and Rankine-Hugoniot jump conditions must be accomplished.  To do
so, the spatial derivative term must be evolved using a Godunov-type
method (e.g. an HLL-type Riemann solver).}

{  In the appendix it is shown that the conservative set of
equations}~\eqref{eq:conservative-vector} {can be discretised in the form of
relation}~\eqref{eq:disc-hll-fvm-2}, {which can be written in a semi-discrete
form as}:

\begin{equation}
\begin{split}
 \frac{ \partial q_a(x_i)  }{ \partial t } = - \frac{ 1 }{\Delta x}
   \Big( &[F_a^{\scriptstyle HLL}]_{i+1/2}^n \\
  - &[F_a^{\scriptstyle HLL}]_{i-1/2}^n  \Big),
\end{split}
\label{semi-discrete}
\end{equation}

\noindent {where} \( \boldsymbol{F}^{\scriptstyle HLL} \) {stands for the
HLL-type Riemann solver approximation for the spatial fluxes (see
appendix).  Using the chain rule on the left hand side of the previous 
equation, it follows that:}

\begin{equation}\label{eq:A-equation}
\begin{split}
\frac{\partial u_a(x_i)}{\partial t} = - \frac{ \mathcal{A}_{ab}(x_i,t_n) }{\Delta x}
   \Big( &[F_a^{\scriptstyle HLL}]_{i+1/2}^n \\
  - &[F_a^{\scriptstyle HLL}]_{i-1/2}^n  \Big).
\end{split}
\end{equation} 

\noindent where $\mathcal{A} = (\partial q / \partial u)^{-1}$. {By applying
a forward-difference formula scheme on the left hand side of  
equation}~(\ref{eq:A-equation}), {we get}

\begin{equation}\label{eq:PVRS}
\begin{split}
{u}_a(x_i,t_{n+1}) = {u}_a(x_i,t_n)-\frac{\Delta t}{\Delta
x}\mathcal{A}_{ab}(x_i,t_n) \Big( [&F_{b}^{\scriptstyle HLL}]_{i+1/2}^n \\
 - &[F_{b}^{\scriptstyle HLL}]_{i-1/2}^n\Big).
\end{split}
\end{equation}

In equation~(\ref{eq:PVRS}), we take a numerical flux approach as in
standard FVM and a finite difference of the time derivative over the
primitive variables $\boldsymbol{u}$. The approximate solution to
the Riemann problem, where Rankine-Hugoniot's condition take place,
is the same as the one presented in the appendix section~\ref{hll}. Furthermore,
the characteristic velocities used in the HLL solver which correspond to
the the eigenvalues of the Jacobian $\partial \boldsymbol{f} / \partial
\boldsymbol{q}$, can be computed either from matrix $M_{ab}$~(\ref{eq:M})
or from \( \partial f_a / \partial q_b \) since
both matrixes are similar~\citep{font1994}. All matrixes and vectors
$(M_{ab}, \mathcal{A}_{ab}, f_a, q_a)$ are computed using a piecewise
reconstruction $\tilde{u}$ of the primitive variables, except for matrix
$A_{ab}$ which is evaluated on the midpoint $x_i$ of the cell \( C_i \).

  It is important to note that in equations~\eqref{eq:azt-qua-eq}
and~\eqref{eq:PVRS} the second term on the right hand side has an implicit
sum over the repeated index \( a \).

  {Note that, although it seems that the PVRS
discretisation}~\eqref{eq:PVRS} {arises directly from discretising
the hybrid quasilinear equation} \( \partial \boldsymbol{u} /
\partial t  + \mathcal{A}\, \partial \boldsymbol{f} / \partial x
= 0\) {--which can be directly obtained by using the chain rule on
equation~}\eqref{eq:conservative-vector}, {it is impossible to obtain the
PVRS discretisation shown in equation~}\eqref{eq:PVRS} {using a standard
conservative FVM as presented in the appendix, and which satisfies the
entropy and Rankine-Hugoniot jump conditions.}

  By using discretisation~(\ref{eq:PVRS}) on a numerical code, it would no
longer be a  concern to recover the primitive variables from the computed
conservative charges; they would instead be solved directly! Therefore,
it would not be necessary to create a module in the code to obtain the
final required solution \( \boldsymbol{u}(x,t) \).  
In general terms, this procedure works out for any kind
of conservative system in which $\boldsymbol{q}(\boldsymbol{u}(x,t))$
and $\boldsymbol{f}(\boldsymbol{u}(x,t))$ are at least given at some
initial time.

  {The time step evolution of the discretisation}~\eqref{eq:PVRS}
{that we use for our numerical simulations is given by the Method of
Lines (MoL):}

\begin{equation}
  \frac{ \partial \boldsymbol{u}(x_i) }{ \partial t } = -
    \boldsymbol{L}(\boldsymbol{u}(x_i)), 
\label{lines}
\end{equation}

\noindent where \( \boldsymbol{L}(\boldsymbol{u}(x_i)) \) {is the right
hand side of equation}~\eqref{eq:PVRS} (see e.g. \citep{lora2013}), {which
can be further implemented with a Runge-Kutta integration.}

\section{Convergence test for PVRS in relativistic hydrodynamics}
\label{convergence}

In this section we are going to show how this new method handles the
evolution of a relativistic gas in a particular Riemann problem namely the
\emph{shock tube}~\citep[see e.g.][]{lora2013}. This {relativistic
Sod \protect{\citep{sod}} shock tube problem}
is a standard test that any code must fulfil for its validation. It
has an exact analytical solution for both special relativistic and
non-relativistic hydrodynamics and it is used for comparisons with
numerical methods.

We calculated the numerical solution using PVRS
discretisation~(\ref{eq:PVRS}) with an approximate HLL Riemann solver,
a \emph{minmod} limiter for the reconstruction $\tilde{u}$ and a 4th order
Runge-Kutta Method of Lines (MoL-RK4) for the integration. The problem was
solved in the domain $[0,1]$ with $N=800$ identical grid cells. We made
{three relativistic Sod tests} with the initial discontinuity located at $x=0.5$
and with initial states shown on Table~\ref{table:1}. Furthermore,
we compared the numerical results with the exact solution shown
by~\citep{lora2013}. Also, we have estimated the usual $\textup{L}_1$-norm
error for the following different resolutions: $\Delta x_1 = 1/200$,
$\Delta x_2 = 1/400$, $\Delta x_3 = 1/800$, $\Delta x_4 = 1/1600$,
$\Delta x_5 = 3200$ and $\Delta x_6 = 1/6400$.

\begin{table}
\centering
\begin{tabular}{cccccccc}
\hline 
Test & $p_L$ & $v_L$ & $n_L$ & $p_R$ & $v_R$ & $n_L$ & $\kappa$ \\
\hline 
1 & 1.0 & 0.0 & 1.0 & 0.1 & 0.0 & 0.125 & 4/3 \\
2 & 13.33 & 0.0 & 10.00 & 0.1 & 0.0 & 1.0 & 4/3 \\
3 & 1000 & 0.0 & 1.0 & 0.01 & 0.0 & 1.0 & 5/3\\
\hline 
\end{tabular}
\caption{\label{table:1} Initial parameters used for the {relativistic 
Sod \protect{\citep{sod}} shock tube tests described in the article. $\kappa$ stands for the polytropic
index.}}
\end{table}

The time-step condition used in this method is different from 
the commonly used by many authors~\citep[cf.][]{delzanna2002}.  A general
CFL-condition applied to this numerical scheme was constructed by us and
used in the set of examples presented.  The exact condition and its
derivation is a subject beyond the scope of this article and will be 
published elsewhere\footnote{For practical purposes, the time step interval
can be chosen as a sufficiently smaller number than the corresponding
CFL condition (cf. equation~\eqref{courant}).}.  For the examples presented
below, we have chosen a fixed time step for each simulation.

\subsubsection{Test 1: Weak relativistic blast wave}

\begin{figure}
		\centering
		\includegraphics[scale=1.0]{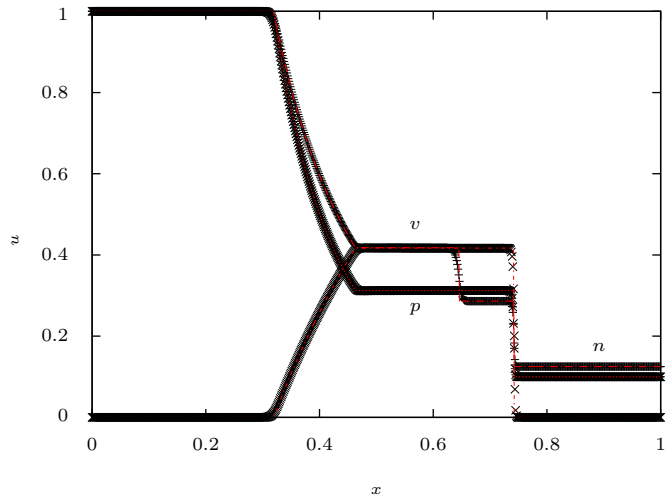}
		\caption{\label{fig:case1} \textit{Test 1}. The figure
		shows the result of the simulation of a weak
		relativistic {(Sod shock tube)} blast wave explosion 
		at t=0.35 for the
		particle number density \( n \), pressure \(p \) and
		velocity \( v \). {The time step used for the simulation was
		0.001.}}
		\end{figure}

The first test corresponds to a lowly relativistic blast wave
explosion. The results can be seen in Figure~\ref{fig:case1}, where
we compare the numerical solution (points) with the exact solution
(lines). It is clear that for both, smooth parts and discontinuities,
the numerical solution converges quite well to the exact one.

\subsubsection{Test 2: Mildly relativistic blast wave}

\begin{figure}
		\centering
		\includegraphics[scale=1.0]{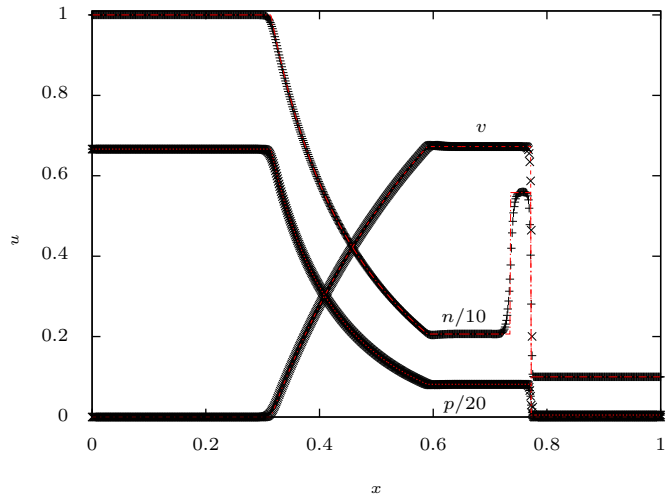}
		\caption{\label{fig:case13} \textit{Test 2}. The figure
		shows the result of a mildly relativistic {(Sod shock
		tube)} blast wave explosion at t=0.35  for particle number
		density \( n \), pressure \(p \) and velocity \( v \). {The
		time step used for the simulation was 0.001.}}
		\end{figure}

The second test corresponds to a mildly relativistic blast wave explosion.
The results can be seen in Figure~\ref{fig:case13}, where we compare
the numerical solution (points) with the exact one (lines). The
importance of this test is to see if, with a relative high difference
in pressure between both states, the numerical method is capable of solving
the density function at the contact discontinuity.

\subsubsection{Test 3: Strong relativistic blast wave}

\begin{figure}
		\centering
		\includegraphics[scale=1.0]{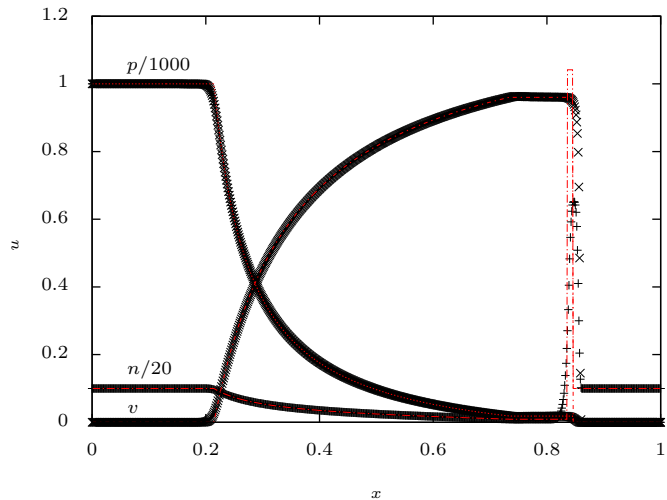}
		\caption{\label{fig:case1000} \textit{Test 3}.
		The figure
		shows the result of a strong
		relativistic {(Sod shock tube)} blast wave explosion 
		at t=0.35  for particle number density \( n \), 
		pressure \(p \) and velocity \( v \).  {The time step used
		for the simulation was 0.0001.}}
		\end{figure}

Finally, the last test corresponds to a strongly relativistic blast
wave explosion. In this case, the density discontinuity is produced by a 
a 5 orders of magnitude difference between right and left initial
detonation pressure, creating a thin shell which numerically is 
harder to resolve at low resolutions. However, with a relatively small number
of cells and a weak variable reconstruction, the results shown on
Figure~\ref{fig:case1000} are as good as the ones obtained by other codes
\citep[cf.][]{delzanna2002,lora2015}.

\subsection{Error estimates}

We have calculated the error of each test using the traditional
L$_1-\text{norm}$ value.  The convergence order of this test is given
by $\log(error_i / error_{i-1}) / log(1/2)$, where $error_j$ is the
L$_1-\text{norm}$ of the $\Delta	x_j$ resolution. As we can
see from Table~\ref{table:2}, the error decreases when the resolution
increases, as expected. Also, we obtain first order convergence for all
test in at least one resolution. Additionally, we made an experiment
following~\cite{yousaf2015} of a static Gaussian curve in order to
estimate the order of convergence of a smooth static profile which,
for this case, reaches a convergence value of about 2 in all the tested
resolutions for a {fixed time step of 0.01}. As expected, this means that 
the important error of the
{relativistic Sod shock tube test} relays on the discontinuities. This
is the reason as to why we consider that taking the L$_1-\text{norm}$
is not a clear indicator of the ``real'' error at the shock waves, so we
propose a more relevant useful visual interpretation of this estimation
as follows.

\begin{figure*}
\centering
\includegraphics[scale=1.0]{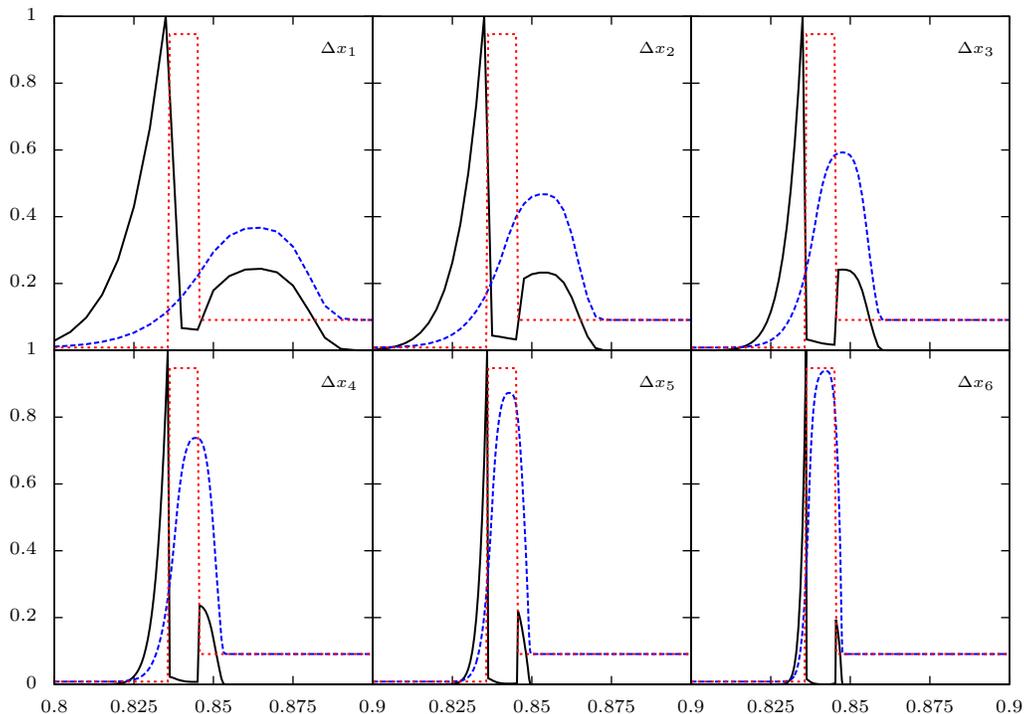}
\caption{\label{fig:error} Comparison between the exact (red dashed-line)
and the numerical solution (blue dashed-line) of the contact discontinuity
in density for the Test~3 vs. the fluctuation $|u_\text{num} -
u_\text{exact}|/u_\text{exact}$ (black continuous line) at each point,
for all the tested resolutions. Note that as the resolution increases,
the width of the fluctuation decreases, showing the convergence in a
straightforward manner.}
\end{figure*}

In Figure~\ref{fig:error} we show both exact (red dashed-line) and
numerical (blue dashed-line) solution vs.  the fluctuation $|u_\text{num}
- u_\text{exact}|/u_\text{exact}$ at each point (black line), for the
density in Test~3 at every resolution. We can see how the Full Width at
Half Maximum (FWHM) of the fluctuation tends to zero as the resolution
increases.  Working with the fluctuation of the numerical solution about
the exact solution is a much better way to easily see the convergence
of a numerical method, rather than the traditional L\(_1\)-norm for
which smoothing of the errors can be wrongly interpreted as a positive
convergence test.

\begin{table*}
\centering
\begin{tabular}{ccccccccccc}
\hline
& \multicolumn{5}{c}{Error} & \multicolumn{5}{c}{Order of Convergence} \\ \hline 
Resolution & & Test 1 & Test 2 & Test 3 & Smooth & & Test 1 & Test 2 & Test 3 & Smooth\\
\hline
$\Delta x_1$ & & 3.83e-3 & 9.00e-2 & 1.93e-1 & 4.74e-4 & & - & - & - & - \\
\hline
$\Delta x_2$ & & 2.12e-3 & 5.04e-2 & 1.60e-1 & 1.28e-4 & & 0.85 & 0.84 & 0.27 & 1.89 \\
\hline
$\Delta x_3$ & & 1.21e-3 & 2.61e-2 & 1.21e-1 & 0.34e-4 & & 0.81 & 0.95 & 0.40 & 1.91 \\
\hline
$\Delta x_4$ & & 6.68e-4 & 1.51e-2 & 8.03e-2 & 0.09e-4 & & 0.85 & 0.79 & 0.59 & 1.92 \\
\hline
$\Delta x_5$ & & 4.01e-4 & 1.02e-2 & 4.56e-2 & 0.02e-4 & & 0.74 & 0.57 & 0.81 & 2.17 \\
\hline
$\Delta x_6$ & & 2.22e-4 & 5.07e-3 & 2.62e-2 & -       & & 0.83 & 1.00 & 1.05 & - \\
\hline
\end{tabular}
\caption{\label{table:2}The L$_1$-norm for the error in the numerical density
for the \emph{minmod} limiter with different numerical resolutions.The L$_1$-norm 
is computed for all shock-tube and Gaussian tests at time $t = 0.35$. We also 
show the order of convergence between different resolutions. Since
the error decreases when the resolution increases, the PVRS constructed in
the article is stable and converges to the exact solution.}
\end{table*}

\section{Discussion}
\label{discussion}

  In this article we have developed a new numerical algorithm to solve
any set of coupled differential conservative equations for which the
primitive variable vector \( \boldsymbol{u} \) is directly obtained.
This is a forward step in numerical methods, since it avoids any
intermediate step reconstruction of the primitive variable vector from a
previously obtained charge vector \( \boldsymbol{q} \) at all points or
cells in space at each time.  In principle, this means that numerical
codes can be written in a more direct form.  Also,  depending on the
nature of the physical problem to solve, the computational time may be
reduced with this technique.

  For practical purposes, we always had in mind special relativistic
hydrodynamical problems and for this reason the specific techniques used
throughout the article deal with hydrodynamical shock capturing schemes.
We demonstrated in the article that the Primitive Variable Recovery
Scheme (PVRS) showed good convergence for three shock-tube and one
Gaussian tests.  { Further explorations in other directions, such as a
non-static Gaussian test \protect{\citep[e.g.][]{rezzolla}} need to be investigated.  We will explore more
details in future works.}

  The PVRS presented in this article can be implemented straightforward
to any standard hydrodynamical code that already uses HLL Riemann solvers
given by equation~\eqref{eq:PVRS}. 

  {In summary, the PVRS is a numerical maneuver to circumvent the
embroiling construction of the primitive vector once the charge vector
is obtained from any standard procedure used to solve a set of coupled
conservative equations in physical systems.}

  We are constructing a GNU Public Licensed (GPL) free software 
(\url{http://www.gnu.org}) called ``\emph{aztekas}''
(\url{http://www.aztekas.org}) that deals with
relativistic hydrodynamics using this PVRS technique.

\section{Acknowledgements}
\label{acknowledgements}
  This work was supported by DGAPA-UNAM and CONACyT grants (PAPIIT
IN112616 and CB-2014-1~\#240512). AAO, SM and DO acknowledge economic
support from CONACyT (788898, 26344, 255602). 


\bibliographystyle{vancouver}
\bibliography{numerical}


\appendix

\section{Traditional approach for numerically solving conservative equations}
\label{traditional}

  In this appendix, we deal with traditional well known methods for solving
conservative equations.  Our intention is to briefly introduce the less
versed reader to this topics using Einstein's summation convention.

A system of $m$ conservative equations in one dimension is usually written
as:

\begin{equation}\label{eq:conservative}
\frac{\partial q_a}{\partial t} + \frac{\partial f_a(q_1,...,q_m)}{\partial x} = 0,
\end{equation}

\noindent where the subindex $a$ takes values from 1 to $m$,
$\boldsymbol{q} := \boldsymbol{q}(\boldsymbol{u}(x,t))$ is the vector of \emph{conservative charges} and
$\boldsymbol{f}:=\boldsymbol{f}(\boldsymbol{q}(\boldsymbol{u}(x,t)))$ is the corresponding \emph{flux} vector along the $x$
axis at a given time \( t \).  The vector $\boldsymbol{u}$ corresponds
to the \emph{primitive variables} for which its number of entries and
functional form of $\boldsymbol{q}(\boldsymbol{u})$ depends on the
particular problem to solve\footnote{From
this point onwards, we are going to use $\boldsymbol{f}(x,t)$ instead of
the cumbersome notation $\boldsymbol{f}(\boldsymbol{q}(\boldsymbol{u}(x,t)))$,
bearing in mind that both, charges and flux vectors, depend on the
primitive variables $\boldsymbol{u}(x,t)$.}. As it is shown in section~\ref{primitive},
the fluxes also have an explicit dependence on the primitive variables
but are usually written in terms of the conservative charges.

  We can rewrite equation~(\ref{eq:conservative}) in the
\emph{quasilinear} following form 

\begin{equation}\label{eq:quasilinear}
  \frac{\partial q_a}{\partial t} + J_{ab} \frac{\partial q_b}{\partial x}
    = 0,
\end{equation}

\noindent where $J_{ab}$ is the \emph{Jacobian} matrix of
$\boldsymbol{f}(\boldsymbol{q})$. From now on, we use Einstein implicit sum convention
over two repeated subindexes contained in the set $\lbrace a,b,c,d
\rbrace$.  If the Jacobian matrix satisfies the conditions of having 
real eigenvalues and a set of independent eigenvectors, then we say that the
system~(\ref{eq:conservative}) is 
\emph{hyperbolic}~\citep[see e.g.][]{leveque2002}.

In the linear cases (when $\boldsymbol{f}$ is a linear function of $\boldsymbol{q}$), there exists
an analytical solution for (\ref{eq:conservative}), but many physical
cases give rise to nonlinear conservative systems which are required to
be solved using numerical methods.

In the following subsections we briefly mention two of the main 
numerical methods used to solve 1D conservative systems such as the one
written in equation~\eqref{eq:conservative}.

\subsection{Finite differences approach}
\label{finite-diff}

The finite differences method (FDM) is one of the most useful and
simple numerical methods for solving ordinary and partial differential
equations. It consists of an approximation of the derivatives of
fluxes and charges based on approximations of their values on sufficiently
small intervals of space and time. The space is divided in a grid of $N$
centred points spaced by equal length $\Delta x$ intervals in which
the equation is evaluated.

Using  Taylor expansions of the involved quantities, it is possible to
work out the finite difference form of equation~(\ref{eq:conservative})
to find the value of $\boldsymbol{q}$ in all the grid at time $t+\Delta
t =: t_{n+1}$ based on its value at $t =: t_{n}$:

\begin{equation}\label{eq:conserv-finite-diff}
\begin{split}
q_a(x_i,t_{n+1}) = q_a(x_i,t_n) - \frac{\Delta t}{2\Delta x}
\big[&f_a(x_{i+1},t_n) \\- &f_a(x_{i-1},t_n) \big],
\end{split}
\end{equation}

\noindent where $x_i$ is the $i$-th point on the grid. This is the
Forward Time Central Space (FTCS) Euler method~\citep{leveque2002}\footnote{In
equation~\eqref{eq:conserv-finite-diff}, the derivative \( \partial f_a
/ \partial x \) at a given time \( t_n \) was written using a central
approximation value given by \( ( f_a(x_{i+1}) - f_a(x_{i-1}) ) / (
2 \Delta x ) \).  For the left and right boundary points this derivative
can be written using a right or left derivative approximation given by: \(
( f_a(x_1)  - f_a(x_0) ) / \Delta x  \) and \( ( f_a(x_{N-1})  - f_a(x_N)
) / \Delta x \) respectively.}. Unfortunately,
discretisation (\ref{eq:conserv-finite-diff}) leads to numerical unstable
solutions~\cite{vonneumann1950}. To overcome this problem, many higher order methods
have been developed and successfully implemented over
time~\citep{laney1998}.

When a second-order finite differences approximation method is
used, additional source \emph{artificial viscosity} terms appear
in~\eqref{eq:conserv-finite-diff}.  Those additional terms are either
due to the second derivative approximation in Taylor series or to
second differences approximation of the first derivatives~\citep[see
e.g.][]{laney1998}. The artificial viscosity name was given by von
Neumann~\citep{vonneumann1950} since it resembles the viscosity term
of the Navier-Stokes equation, but has nothing to do with any physical
viscosity.

The general form of the artificial viscosity can be written
as~\citep{laney1998}:

\begin{equation}\label{eq:art-vis}
\begin{split}
q_a(x_i,t_{n+1}) = &q_a(x_i,t_n) \\
- \frac{\Delta t}{2\Delta x} \big[&f_a(x_{i+1},t_n) - f_a(x_{i-1},t_n) \big] \\
+ \frac{\Delta t}{2\Delta x} \big[ &\epsilon_a^+ \Delta q_a^+(x_i,t_n) - \epsilon_a^- \Delta q_a^-(x_i,t_n)\big],
\end{split}
\end{equation}

\noindent where $\epsilon_a^{\pm}$ are the \emph{coefficients
of second-order explicit artificial viscosity} and $\Delta
q_a^{\pm}(x_i,t_n) = \pm q(x_{i\pm 1},t_n) \mp q(x_i,t_n)$. The choice
$\epsilon_a^{\pm} = 2 \Delta x / \Delta t$ simplifies the above equation
to:

\begin{equation}
\begin{split}
q_a(x_i,t_{n+1}) = \frac{1}{2} \big(&q_a(x_{i+1},t_n) + q_a(x_{i-1},t_n) \big) \\
- \frac{\Delta t}{2 \Delta x} \big[ &f_a(x_{i+1},t_n) - f_a(x_{i-1},t_n) \big],
\end{split}
\end{equation}

\noindent which is known as the \emph{Lax-Friedrich method}. Other
second-order-two-step methods, such as the \emph{Lax-Wendroff method},
have been developed and successfully implemented in many numerical codes.

\begin{figure}
\centering
\includegraphics[scale=1.0]{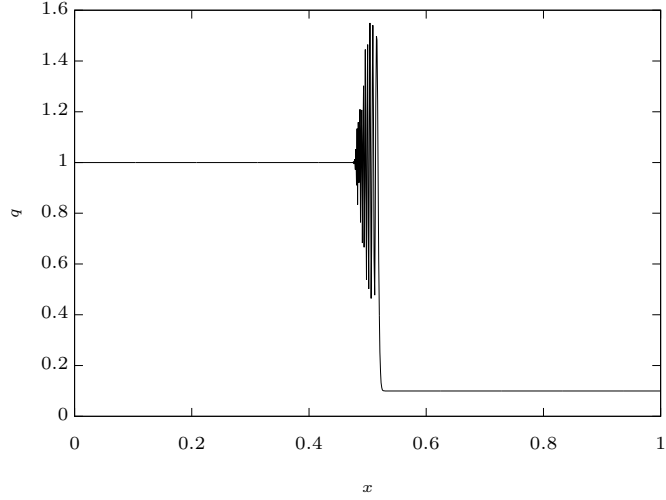}
\caption{\label{fig:gibbs}The graph shows the numerical
solution of the advection equation: $\partial_t q + \partial_x q =0$, 
using exclusively the MacCormack method. The solution shows non-physical
oscillations in a finite-jump discontinuity due to the Gibbs phenomenon.
At later times, the oscillations grow breaking even more the expected
solution.  The graph was constructed using the initial conditions
of $q = 1.0$ if $x < 0.5$ and $q = 0.125$ elsewhere at a fixed time \( t = 0.03 \).}
\end{figure}

 One such favourite two-step method was proposed by~\citep{maccormack1972}.
It makes a \textit{forward-prediction} of $\boldsymbol{q}$ and with it, a
\textit{backward-correction}:

\begin{equation}\label{predictor}
\begin{split}
\tilde{q}_a(x_i,t_n) := q_a(x_i,t_n) - \frac{\Delta t}{\Delta x} \big[&f_a(x_{i+1},t_n) \\
- &f_a(x_{i},t_n)  \big],
\end{split}
\end{equation} 
\begin{equation}\label{corrector}
\begin{split}
q_a(x_i,t_{n+1}) = \frac{1}{2} \bigg \lbrace &q_a(x_i,t_n) + \tilde{q}_a(x_i,t_n) \\
- \frac{\Delta t}{\Delta x}\big[ &\tilde{f}_a(x_{i},t_n) - \tilde{f}_a(x_{i-1},t_n) \big] \bigg \rbrace .
\end{split}
\end{equation}

\noindent where $\tilde{\boldsymbol{f}} :=
\boldsymbol{f}(\tilde{\boldsymbol{q}})$.  This method has been proved
to be consistent, convergent and stable which is the requirement
for any numerical method used in a computational code. Nevertheless,
in discontinuities and regions with high pressure gradients, such as
regions with shock-waves, this algorithm introduces a dispersive error
called the Gibbs phenomenon, which consists on the presence of large
spurious oscillations near the finite-jump, such as the example shown
in Figure~\ref{fig:gibbs}.

To solve this problem, it is common to apply a \emph{corrective diffusion}
in the regions where the non-physical oscillations appear. The correction
presented by \citep{book1975} is

\begin{equation}
\begin{split}
q_a^*(x_i,t_n) = q_a(x_i,t_n) &+ \eta [q_a(x_{i+1},t_n) \\
 &- 2q_a(x_i,t_n) + q_a(x_{i-1},t_n)],
\end{split}
\end{equation}

\noindent where $\eta$ is the \emph{antidiffusion coefficient} at
space-time points $x_i$ and $t_n$:

\begin{equation}\label{antidiffusion}
\eta = \left\lbrace
\begin{array}{ll}
\eta_0 \leq 1/4, & \textup{if} \quad (\Delta q_a^+)(\Delta q_a^-) < 0, \\
0, & \textup{if} \quad (\Delta q_a^+)(\Delta q_a^-) > 0.
\end{array}
\right.
\end{equation}


\subsection{Finite volume approach}
\label{finite-vol}

A more natural way of obtaining the discretisation form 
of~(\ref{eq:conservative}) is the Finite Volume Method (FVM)
which is based on a subdivision of  the spatial domain into intervals
(also called \emph{control volumes} or \emph{grid cells}) $C_i :=
[x_{i-1/2},x_{i+1/2}]$. The integration of~(\ref{eq:conservative}) 
over $C_i$ between times $t_n$ and $t_{n+1}$ yields:

\begin{equation}\label{integral-form}
\int_{t_n}^{t_{n+1}} \int_{C_i} \bigg[ \frac{\partial q_a(x,t)}{\partial
t} + \frac{\partial f_a(x,t)}{\partial x} \bigg] \textup{d}x \, \textup{d}t
= 0.
\end{equation} 

\noindent The integral of $\partial_t q$ over time
and the integral of $\partial_x f$ over space can be solved exactly and so,
the next integral form of the previous equation is found:

\begin{equation}\label{integral-form-2}
\begin{split}
&\int_{C_i} (q_a(x,t_{n+1}) - q_a(x,t_n)) \textup{d}x \\ +
&\int_{t_n}^{t_{n+1}} (f_a(x_{i+1/2},t) - f_a(x_{i-1/2},t)) \textup{d}t
= 0.
\end{split}
\end{equation}

 At this point, both integrals in the previous equation cannot be
integrated unless we have the exact form of $q$, which is precisely the
solution to the problem.  In order to overcome this, we define each 
integration as a new numerical vector in the following form:

\begin{equation}\label{ave-charge}
[Q_a]_i^n = \frac{1}{\Delta x}\int_{C_i}q_a(x,t_n)\textup{d}x,
\end{equation}
\begin{equation}\label{ave-flux}
[F_a]_{i \pm 1/2}^n = \frac{1}{\Delta t} \int_{t_n}^{t_{n+1}}f_a(x_{i \pm 1/2},t)\textup{d}t,
\end{equation}

\noindent where $[Q_a]_i^n$\footnote{From now on, the square brackets notation
$[ \ ]$ around any numerical function is used to denote the 
corresponding (space or time) average related 
to that specific numerical function.} is the \emph{average charge
vector} of $q$ over $C_i$ at time $t_n$ and $[F_a]_{i \pm 1/2}^n$ is
the \emph{average flux vector} across the boundaries of $C_i$.

If $\boldsymbol{q}(\boldsymbol{u}(x,t))$ is a smooth function, then the
integral (\ref{ave-charge}) agrees with the value of $\boldsymbol{q}$
at the midpoint of the interval to $\mathcal{O}(\Delta x^2)$
\citep{leveque2002}.

The indexes outside the square bracket do not denote the spatial and
time evaluation of the average vector, they are just labels that refer
to the time and grid positions of the corresponding numerical values.

Substituting the definitions (\ref{ave-charge}) and (\ref{ave-flux})
in equation (\ref{integral-form-2}) we obtain the main discretisation for
the finite volume scheme usually presented in the literature
\citep[cf.][]{leveque2002}:

\begin{equation}\label{fvm-disc}
[Q_a]_i^{n+1} = [Q_a]_i^n - \frac{\Delta t}{\Delta x} \big( [F_a]_{i+1/2}^n - [F_a]_{i-1/2}^n \big).
\end{equation}

Equation (\ref{fvm-disc}) is a numerical recipe of how to compute the
mean value $[Q_a]_i^{n+1}$ using the average flux and charge values 
one time-step backwards for each grid cell $C_i$. This
discretisation has the same exact form as~(\ref{eq:conservative}) except for
the choice of the values~(\ref{ave-charge}) and~(\ref{ave-flux}).

The advantage of this method over any finite difference scheme is that
the conservative nature of the system is preserved, even across strong
discontinuities such as shock waves. This is the reason as to why 
a finite volume scheme is often used when dealing with the physics of high
energy flows where discontinuities may appear.

\subsubsection{Numerical flux}

The flux $\boldsymbol{f}$ at (\ref{ave-flux}) depends on the value of $q$ at
every time. This is why it is impossible to integrate the
average flux. Somehow, we have to find a good approximation for
this integral. Moreover, the flux $\boldsymbol{f}$ inside the integral is evaluated
on the boundaries $x_{i \pm 1/2}$ of the grid cell  which, numerically
speaking, has no sense because we can only approximate the values of the
average charges on the midpoint of the finite volume\footnote{This set of
midpoints can be ''safely'' considered the ones used in the finite 
difference mesh mentioned in section~\ref{finite-diff}.}.

One way to approximate $[F_a]_{i \pm 1/2}^n$ is to assume that it
can be obtained as a function of the cell average values of $\boldsymbol{q}$ on
either side of the interface $x_{i \pm 1/2}$, i.e., $[Q_a]_{i \pm 1}^n$
and $[Q_a]_i^n$:

\begin{equation}\label{num-flux}
[F_a]_{i \pm 1/2}^n = \mathcal{F}_a \big( [Q_a]_{i \pm 1}^n,[Q_a]_i^n \big).
\end{equation}

\noindent The previous result is expected since in a hyperbolic problem
the information of how $\boldsymbol{q}$ change on every cell propagates
at a finite characteristic speed~\citep[see e.g.][]{laney1998,landau1987}.  The function $\mathcal{F}_a$ can be thought as a \emph{numerical flux function} for
which its functional form will depend on the problem or the particular
numerical scheme used to solve it.

 Substitution of equation~(\ref{num-flux}) into~(\ref{fvm-disc})
yields:

\begin{equation}\label{fvm-disc-2}
\begin{split}
[Q_a]_i^{n+1} = [Q_a]_i^n - \frac{\Delta t}{\Delta x} \big[ &\mathcal{F}_a \big( [Q_a]_{i+1}^n,[Q_a]_i^n \big) \\
- &\mathcal{F}_a \big( [Q_a]_{i-1}^n,[Q_a]_i^n \big) \big].
\end{split}
\end{equation}

\noindent The numerical flux function is then determined by the evolution of
the solution in each interface. A good first guess for the function \(
\mathcal{F}_a \) is to relate it to the corresponding average flux function
of a local (for each cell) Riemann problem~\citep{lora2013} with two
constant states on each side of the boundary.

In order to obtain an accurate numerical flux function, is important to
study the behaviour of the solution based on the form and properties of
the governing equation at these particular initial conditions.

\subsubsection{Riemann problem}

Let us now consider a single conservative equation
(i.e. relation~\eqref{eq:conservative} with \( a = 1 \) only) in which the
flux is written as $f(q) = \tilde{u}q$ where $\tilde{u}$ is a constant
value:

\begin{equation}\label{eq:conserv-1d} 
  \frac{\partial q}{\partial
    t} + \tilde{u}\frac{\partial q}{\partial x} = 0. 
\end{equation}

\noindent This is the advection equation in which $\tilde{u}$ corresponds to
the propagation velocity of $q$. Note that, since $f'(q) = \tilde{u}$,
equation~(\ref{eq:conserv-1d}) is also its own quasilinear version.

 The function $q(x,t) = \tilde{q}(x-\tilde{u}t)$
satisfies equation~(\ref{eq:conserv-1d}) for any function
$\tilde{q}$. However, it is more useful for us to describe the problem
observing the behaviour of the solution $q$ along \emph{characteristic
curves} in the $t-x$ plane. To do so, we perform the time derivative 
of $\boldsymbol{q}(X(t),t)$ and equate the result to zero, i.e.: 

\begin{equation}\label{time-derivative}
  \frac{\textup{d}}{\textup{d}t} q(X(t),t) = \frac{\partial q}{\partial t}
    + X'(t)\frac{\partial q}{\partial x} = 0.
\end{equation}

\noindent Direct comparison of the above equation
with~(\ref{eq:conserv-1d}), means that 
that the solution $q(X(t),t)$ is constant all along the ray $X(t) =
x_0 + \tilde{u}t$, where $x_0$ is some initial value. In the most general
case, the set of all rays $X(t)$
are called the \emph{characteristics} of the equation.

If we consider the particular case in which the initial conditions of
the problem consists on two constant states

\begin{equation}\label{solution-riemann}
q(x,0) = \left\lbrace
\begin{array}{ll}
q_l, & \textup{if} \quad x < 0,  \\
q_r, & \textup{if} \quad x > 0,
\end{array}
\right.
\end{equation}

\noindent where $q_l$ and $q_r$ are the left and right states
respectively, the characteristics $X(t)$ of~(\ref{eq:conserv-1d}) are
then rays with slope $\tilde{u}$ in the $t-x$ plane. With this, the solution
can be written as

\begin{equation}\label{solution-riemann2}
q(x,t) = \left\lbrace
\begin{array}{ll}
q_l, & \textup{if} \quad x-\tilde{u}t < 0 \quad \textup{or} \quad x/t < \tilde{u},  \\
q_r, & \textup{if} \quad x-\tilde{u}t > 0 \quad \textup{or} \quad x/t > \tilde{u}.
\end{array}
\right.
\end{equation}

Let us consider now a system of $m$ conservative equations (i.e. \( a = 1,\
2,\ \ldots, m \) in relation~\eqref{eq:conservative}), where $f_a =
A_{ab}q_b$, i.e.:

\begin{equation}\label{linearsys}
\frac{\partial q_a}{\partial t} + A_{ab}\frac{\partial q_b}{\partial x} = 0,
\end{equation}

\noindent where $A_{ab}$ is a constant $m \times m$ matrix and so, the
system of conservative equations is linear. If
$A_{ab}$ is diagonalisable such that:

\begin{equation}\label{diagonalizable}
A_{ab} = R_{ac} \Lambda_{cd} R_{db}^{-1},
\end{equation}

\noindent where $R_{ac}$ is the matrix of eigenvectors, with $r_a^p$
the $p$-th eigenvector, $R_{db}^{-1}$ its inverse
and $\Lambda_{cd} = \textup{diag}(\lambda^1, ..., \lambda^m)$,
for  $\lambda^p$ the $p$-th eigenvalue. If we define the
\emph{characteristic variables} $w_a$ as

\begin{equation}\label{charvar}
  w_a(x,t) := R_{ab}^{-1}q_b(x,t),
\end{equation}

\noindent it is then possible to rewrite equation~\eqref{linearsys} as 
the following system of $m$ advective equations:

\begin{equation}\label{adv-system}
\frac{\partial w_a}{\partial t} + \Lambda_{ab} \frac{\partial
w_b}{\partial x} = 0.
\end{equation}

\noindent In the case of the Riemann problem, the solution for the
$p$-th advective equation is $w_p(x,t) = \tilde{w}_p(x - \lambda^p t,0)$,
and the solution $q_a(x,t)$ is obtained using the definition of $w_a$:

\begin{equation}
q_a(x,t) = R_{ab}\tilde{w}_b(x,t).
\end{equation}

In this way one can think that $q_a$ is a superposition of $m$ waves 
moving with \emph{characteristic velocities} $\lambda^1$, $\lambda^2$,
\ldots and $\lambda^m$, respectively \citep{laney1998}.

Another way to see this is by comparing equation~(\ref{linearsys}) with
the time derivative of $q(X(t),t)$ in~\eqref{time-derivative}. From this, it
follows that the characteristics are curves for which their corresponding
slopes are exactly the eigenvalues of the matrix $A_{ab}$.

In order to obtain a real contribution of one of these waves to the
evolution of a contiguous grid cell, the size of the control volume must
be larger than the distance travelled by the wave, moving at its
characteristic velocity, at a certain fixed time $\Delta t$, i.e.,

\begin{equation}\label{courant}
\lambda \frac{\Delta t}{\Delta x} < 1.
\end{equation}

\noindent The quantity $\lambda \Delta t / \Delta x$ is know as the
\emph{Courant number} and the fulfilment of relation (\ref{courant})
is called \emph{Courant-Friedrich-Levy (CFL) condition}. This is
a convergence requirement for several numerical methods that solve
conservative equations.

The Riemann problem discussed in this subsection, is used to accurately  estimate
the value of the numerical fluxes at the boundaries of two contiguous grid
cells as will be seen in the following section.

\subsubsection{Godunov scheme}

\citep{godunov1959} proposed a numerical scheme for solving conservative
equations and this method can be used in terms of the Riemann problem as
follows.  Consider the single equation~(\ref{eq:conserv-1d}). The algorithm
proposed by Godunov has the following recipe:

\begin{enumerate}
	\item \label{uno} Compute the average values 
	of the charges $q$ at the time
	$t=t_n$ using equation~\eqref{ave-charge} for \( a = 1 \) only:
	\begin{equation}
	[Q]_i^n = \frac{1}{\Delta x} \int_{C_i} q(x,t_n)\textup{d}x.
	\end{equation}
	
	\item Reconstruct from $[Q]_i^n$ a polynomial function
	$\tilde{q}(x,t_n)$ for every value of $x$. The simplest case
	for this is to take a constant function:

	\begin{equation}
	\tilde{q}(x,t_n) := [Q]_i^n \quad \textup{for} \quad x \in C_i.
	\end{equation}
	
	\noindent In practice~\citep[eg.][]{rezzolla2013}, the value
	$[Q]_i^n$ is consider to be $q$ evaluated at the midpoint of
	the grid cell.

	\item Evolve the hyperbolic equation in an exact or approximate
	way by a time $\Delta t$ to obtain $\tilde{q}(x,t_{n+1})$.

	\item Take the average of $\tilde{q}(x,t_{n+1})$ over $C_i$
               to obtain $[Q]_i^{n+1}$.

	\item Go back to the first item on the list and iterate until
	      a final time is reached.

\end{enumerate}

As we discuss above, it is impossible to compute exactly the average flux
$[F]_{i \pm 1/2}^n$ because we do not know the value of $q$ at all times. 
However, if we consider a Riemann problem in the interface $x_{i \pm
1/2}$ between the grid cells $C_i$ and $C_{i \pm 1}$ and apply step~3 
of Godunov's algorithm, we get that $\tilde{q}(x_{i \pm 1/2},t)$ is
constant along the curves that satisfies $(x - x_{i \pm 1/2})/t =
\text{const.}$

In summary, if we denote by $q^{\downarrow}([Q]_i^n , [Q]_{i \pm 1}^n)$ 
the solution to the Riemann problem at $x_{i \pm 1/2}$, the computation of the
average fluxes reduces on computing an integral over a constant function~\citep{leveque2002}.  In this way, the Godunov's algorithm can be 
expressed in terms of average fluxes using the following recipe:

\begin{enumerate}
	\item Solve the Riemann problem in the interfaces $x_{i \pm 1/2}$
	of the $C_i$ grid cell in order to obtain $q^{\downarrow}([Q]_i^n
	,[Q]_{i \pm 1}^n)$.
	
	\item Define $\mathcal{F}([Q]_i^n , [Q]_{i \pm 1}^n) =
	f(q^{\downarrow}([Q]_i^n , [Q]_{i \pm 1}^n))$.

	\item Apply discretisation~(\ref{fvm-disc-2}).
\end{enumerate}

The problem with applying Godunov's scheme on non-linear systems and
considering wave propagation of characteristic waves on all interfaces, 
is that the characteristic velocities
are not constant at all times and also they change values at different
grid cells. For the case of a quasilinear system such as the one
of equation~(\ref{eq:quasilinear}), an approximation has to be
made. Many methods for obtaining an approximate Riemann solution
have been developed and successfully implemented in classical and
relativistic magnetohydrodynamic codes (see e.g.~\citep{miyoshi2005},
~\citep{lora2015}).

\subsubsection{HLL Riemann solver}
\label{hll}

\begin{figure}
\includegraphics[scale=0.25]{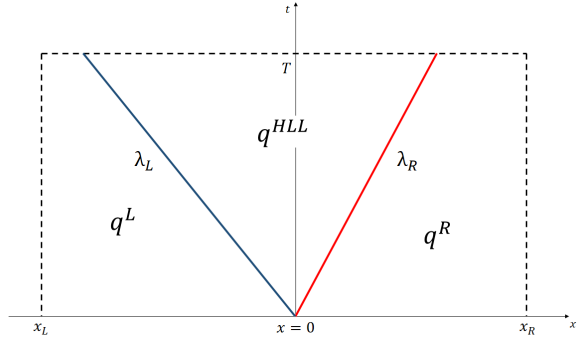}
\caption{\label{hllimage}Space-time grid cell $[x_L,x_R] \times
[0,T]$. The figure shows the evolution of the 1D conservative equation
solution along rays with slope $\lambda_L$ and $\lambda_R$, together
with the intermediate state $q^{HLL}$ generated by the HLL solver.}
\end{figure}

One of the most popular approximate Riemann solvers is the one proposed
by~\citep{HLL1983}. This Godunov's base method considers a
Riemann problem with constant states $\boldsymbol{q}^L$ and $\boldsymbol{q}^R$ on each side of the
interface in a space-time grid cell $[x_L,x_R] \times [0,T]$ as shown
on Figure~\ref{hllimage}.

Instead of following
the solution of all the characteristic variables along their own
characteristic velocities, the idea of the HLL approximation consists on 
considering the larger eigenvalues $\lambda_R$ and $\lambda_L$  moving
across the interface to the right and left respectively.  The region
delimited by these characteristic rays  is denoted by the state
$\boldsymbol{q}^{HLL}$.

Note that, since we are working with a system of \( m \) conservative
equations, \( 2m \)  characteristic rays will emerge from each interface.  
The values $\lambda_L$ and $\lambda_R$ are to be chosen
taking into account all \( 2m \) characteristic  velocities.

The approximate solution to the Riemann problem derived by this scheme
has the following form (see e.g. ~\citep{leveque2002} or~\citep{toro2009}):

\begin{equation}\label{eq:qhll}
q_a(x,t) = \left\lbrace
\begin{array}{ll}
q_a^L, & \textup{if} \quad x/t \leq \lambda_L, \\
q_a{\scriptstyle HLL} & \textup{if} \quad \lambda_L < x/t < \lambda_R, \\
q_a^R, & \textup{if} \quad x/t \geq \lambda_R ,
\end{array}
\right.
\end{equation}

\noindent where 
\begin{equation}\label{eq:qhll2}
q_a^{HLL} = \frac{\lambda_R q_a^R - \lambda_L q_a^L + f_a^L - f_a^R}{\lambda_R - \lambda_L},
\end{equation}

\noindent where $\boldsymbol{f}^{R,L} := \boldsymbol{f}(\boldsymbol{q}^{R,L})$. One can work out the approximate
solution to the flux through the interface by integrating the hyperbolic
equation over the space-time domain outlined in Figure~\ref{hllimage}
and using the Rankine-Hugoniot jump condition at each characteristic ray
($\lambda_{R,L}$). The final result is that~\citep{toro2009}:

\begin{equation}\label{eq:fhll}
f_a^{HLL} = \frac{\lambda_R f_a^L - \lambda_L f_a^R + \lambda_R \lambda_L
(q_a^R - q_a^L)}{\lambda_R - \lambda_L}.
\end{equation}

\noindent Notice that $\boldsymbol{f}^{HLL} \neq
\boldsymbol{f}(\boldsymbol{q}^{HLL})$. The flux~\eqref{eq:fhll}
can be used
along with the Godunov scheme to solve the local Riemann problem of
to contiguous grid cells.

Let us now consider the boundary $x_{i - 1/2}$ between two control
volumes $C_i$ and $C_{i-1}$ and suppose that a constant reconstruction
$\tilde{\boldsymbol{q}}$ from the average values of $\boldsymbol{q}$ has been made. With this, let
$\tilde{{q}_a}^L(x_{i - 1/2},t_n) := [Q_a]_{i-1}^n$ and 
$\tilde{{q}_a}^R(x_{i - 1/2},t_n) := [Q_a]_i^n$ to be 
the reconstruction points that lay at the interface $x_{i - 1/2}$.
Note that these values are
going to be different if a polynomial reconstruction is made.
With this, we can write the numerical flux at $x_{i - 1/2}$ used in the 
Godunov scheme in the following form:

\begin{equation}\label{eq:hllnumflux}
[F_a^{HLL}]_{i - 1/2}^n = \left\lbrace
\begin{array}{llc}
f_a^L(x_{i-1/2},t_n), & \textup{if} & 0 \leq \lambda_L , \\
f_a^{\scriptstyle HLL}(x_{i - 1/2},t_n) & \textup{if} & \lambda_L < 0 < \lambda_R, \\
f_a^R(x_{i-1/2},t_n), & \textup{if} & 0 \geq \lambda_L .
\end{array}
\right .
\end{equation}

\noindent The flux through $x_{i + 1/2}$ is obtained in an
analogous way. So, by substituting these numerical fluxes in the
discretisation~(\ref{fvm-disc-2}), we finally get the numerical
solution for the hyperbolic equation~(\ref{eq:conservative}) in the
finite volume scheme using Godunov's algorithm with a \emph{high
resolution}\citep{marti2003} approximate Riemann HLL solver:

\begin{equation}\label{eq:disc-hll-fvm}
\begin{split}
[Q_a]_i^{n+1} = [Q_a]_i^n - \frac{\Delta t}{\Delta x} \Big( &[F_a^{\scriptstyle HLL}]_{i+1/2}^n \\
- &[F_a^{\scriptstyle HLL}]_{i-1/2}^n  \Big). 
\end{split}
\end{equation}

\noindent A simple way of computing $[Q_a]_i^n$ is by considering that
this average value match the magnitude of $\boldsymbol{q}$ evaluated at
the midpoint of the grid cell $x_i$. If $\boldsymbol{q}(x,t)$ is smooth,
the error introduced by this approximation is of order $\mathcal{O}(\Delta
x^2)$ \citep{leveque2002}.  In other words:

\begin{equation}\label{eq:disc-hll-fvm-2}
\begin{split}
q_a(x_i,t_{n+1}) = q_a(x_i,t_n) - \frac{\Delta t}{\Delta x} \Big( &[F_a^{\scriptstyle HLL}]_{i+1/2}^n \\
- &[F_a^{\scriptstyle HLL}]_{i-1/2}^n  \Big). 
\end{split}
\end{equation}

Many other HLL-type Riemann solvers have been developed
\citep[cf.][]{toro2009} and successfully implemented
\citep[cf.][]{miyoshi2005} but they are beyond the scope of the present
article.

\subsubsection{Limiters}

At  first approximation, the reconstruction of $q$ over the grid cell
was made considering a constant value $[Q]_i^n$ which is taken
as the midpoint value of $q$ of the corresponding control volume $C_i$. 
A better way of improving the precision of the above procedure is by 
considering a piecewise polynomial approximation for this variable.

In the linear case, the reconstruction of $q$ over $C_i$ is given by

\begin{equation}\label{eq:reconst}
\tilde{q}(x,t_n) = q(x_i,t_n) + \sigma_i^n(x - x_i),
\end{equation}

\noindent where $\sigma_i^n$ is the slope of the linear
reconstruction. To use the limiters together with a HLL-type Riemann
solver, all we need to consider are those points of $\tilde{q}$ in each
contiguous grid cells, evaluated at the interfaces $x_{i \pm 1/2}$.  In
this respect, it is not important to do a complete reconstruction of $q$.
The knowledge of \( q \) at the boundaries is sufficient for this
approximation, and so 
the values required to effectively evolve the solution of the hyperbolic
equation over the grid cell $C_i$ are:

\begin{equation}\label{eq:lminus}
\tilde{q}^L (x_{i-1/2},t_n) = q(x_{i-1},t_n) + \frac{1}{2}\sigma_{i-1}^n \Delta x,
\end{equation}
\begin{equation}\label{eq:rminus}
\tilde{q}^R (x_{i-1/2},t_n) = q(x_{i},t_n) - \frac{1}{2}\sigma_{i}^n \Delta x,
\end{equation}
\begin{equation}\label{eq:lplus}
\tilde{q}^L (x_{i+1/2},t_n) = q(x_{i},t_n) + \frac{1}{2}\sigma_{i}^n \Delta x,
\end{equation}
\begin{equation}\label{eq:rplus}
\tilde{q}^R (x_{i+1/2},t_n) = q(x_{i+1},t_n) - \frac{1}{2}\sigma_{i+1}^n \Delta x.
\end{equation}

\noindent Each pair~(\ref{eq:lminus}-\ref{eq:rminus}) and
(\ref{eq:lplus}-\ref{eq:rplus}), constitute a Riemann problem to be solved
at the interface $x_{i-1/2}$ and $x_{i+1/2}$, respectively. The polynomial
reconstruction are useful to accurate capture discontinuities such as 
shock-waves.  Equations~\eqref{eq:lminus}-\eqref{eq:rplus} are also valid for
each component of the vector \( \boldsymbol{q} \) when a coupled system of
conservative equations is required.

The usual way of computing $\sigma$ is by considering some useful
function based on finite derivatives of $q$ over $C_i$. The most
used but dissipative reconstruction (also called \emph{limiter}
\citep{leveque2002}) is the \emph{minmod limiter} (MM) introduced
by~\citep{roe1986}: 

\begin{equation}\label{eq:sigmin}
  \sigma_i^n = \textup{minmod}(m_{i - 1/2},m_{i+1/2}),
\end{equation}

\noindent where the function $m_{i \pm 1/2}$ is the average slope 
(or the finite derivative) of $q$ centred at $x_{i \pm 1/2}$:

\begin{equation}\label{eq:cent-deriv1}
m_{i+1/2} = \frac{q(x_{i+1},t_n)-q(x_i,t_n)}{x_{i+1}-x_i},
\end{equation}

\begin{equation}\label{eq:cent-deriv2}
m_{i-1/2} = \frac{q(x_i,t_n)-q(x_{i-1},t_n)}{x_i - x_{i-1}}.
\end{equation}

\noindent The \emph{minmod} function of two values $a$ and $b$
stands for:

\begin{equation}\label{eq:minmod}
\textup{minmod}(a,b) := \left\lbrace
\begin{array}{ccc}
0, & \textup{if} & ab \leq 0, \\
a, & \textup{if} & |a| < |b|, \\
b, & \textup{if} & |b| < |a|.
\end{array} 
\right.
\end{equation}

This limiter has been successfully implemented in the case of relativistic
hydrodynamics~\citep[cf.][]{lora2015,rezzolla2013}.

The monotonic centred limiter \emph{MC}, proposed by \citep{vanleer1977},
has less dissipation than minmod near discontinuities, but has been
proved to create spurious oscillations in the strong shock cases
\citep{lora2015}. Nevertheless, it produces relatively well damped 
solutions that capture not too strong shock waves.
The slope $\sigma$ is written as in~(\ref{eq:sigmin}) 
but the MC function has the following form:

\begin{equation}\label{eq:mc}
\textup{MC}(a,b) := \left\lbrace
\begin{array}{ccccc}
0, & \textup{if} & ab \leq 0, & & \\
2a, & \textup{if} & |a| < |b| & \textup{and} & 2|a| < |c|, \\
2b, & \textup{if} & |b| < |a| & \textup{and} & 2|b| < |c|, \\
c, & \textup{if} & |c| < 2|a| & \textup{and} & |c| < 2|b|, \\
\end{array}
\right.
\end{equation} 

\noindent where $c := (a+b)/2$.

Another piecewise linear reconstruction is the \emph{superbee} limiter,
also proposed by~\citep{roe1986}. This one has a better shock-wave
capture than the previous scheme as shown in Figure~\ref{fig:reconst}, where 
comparisons of the superbee limiter with the previous ones and with
the piecewise constant reconstruction~(\emph{godunov}) is made. For this slope,
the function is slightly more complicated than the previous ones and is
given by:

\begin{figure}
\centering
\includegraphics[scale=1.0]{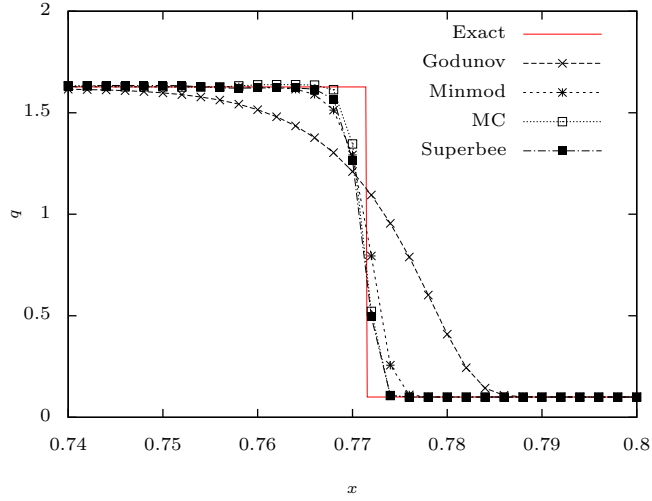}
\caption{\label{fig:reconst} Comparison between the
piecewise linear reconstructions (\textit{minmod},
\textit{MC} and \textit{superbee}) with the piecewise
constant one (\textit{godunov}). As the
complexity of the algorithm grows the shock capture is
better, as it is shown in the figure by the superbee simulation. 
The graph shows the quantity $q$ corresponding to the pressure as a
function of the position at a fixed time $t=0.35$ for a particular
Riemann problem in a {relativistic Sod shock tube} that evolves
from the initial value $t=0$ in such a way that, at this time, $p=1.69$
for $x$ \textless $  0.77$ and $p=0.1$ for $x \geq 0.77$.
}
\end{figure}

\begin{equation}\label{eq:superbee}
\sigma = \textup{maxmod} \left( \sigma_i^{n(1)}, \sigma_i^{n(2)} \right),
\end{equation}

\noindent where
\begin{equation}\label{eq:sup-sig-1}
\sigma_i^{n(1)} = \textup{minmod}\left( m_{i+1/2},2m_{i-1/2} \right),
\end{equation}
\begin{equation}\label{eq:sup-sig-2}
\sigma_i^{n(2)} = \textup{minmod}\left( 2m_{i+1/2},m_{i-1/2} \right),
\end{equation}

\noindent and 
\begin{equation}\label{eq:maxmod}
\textup{maxmod}(a,b) := \left\lbrace
\begin{array}{ccc}
0, & \textup{if} & ab \leq 0, \\
a, & \textup{if} & |b| < |a|, \\
b, & \textup{if} & |a| < |b|.
\end{array}
\right.
\end{equation}

\citep{colella1984} developed a piecewise parabolic reconstruction (PPM), that have been successfully used by many authors in both relativistic~\citep[cf.][]{lora2015} and non-relativistic hydrodynamics~\citep[cf.][]{amik2007} but for the purposes of this paper, it will not be considered.

\end{document}